\documentclass[aps,graphicx,twocolumn,epsfig]{revtex4}

\newcommand{\beq}{\begin{equation}}
\newcommand{\eeq}{\end{equation}}
\newcommand{\beqn}{\begin{eqnarray}}
\newcommand{\eeqn}{\end{eqnarray}}

\usepackage{epsfig}
\begin{document}

\author{L. Viverit$^{1,2}$}
\author{C. Menotti$^{1}$}
\author{T. Calarco$^{3,4}$}
\author{A. Smerzi$^{1}$}
\affiliation{$^1$CRS BEC-INFM and Dipartimento di Fisica,
Universit\`a di Trento, I-38050 Povo, Italy\\
$^2$Dipartimento di Fisica, Universit\`a di Milano,
via Celoria 16, I-20122 Milan, Italy\\
$^3$European Centre for Theoretical Studies in Nuclear Physics and
Related Areas, I-38050 Villazzano (TN, Italy)\\
$^4$Institut f\"ur Theoretische Physik, Universit\"at Innsbruck,
A-6020 Innsbruck (Austria)}

\title{Efficient and robust initialization of a qubit register
with fermionic atoms}

\begin{abstract}
We show that fermionic atoms have crucial advantages over bosonic
atoms in terms of loading in optical lattices for use as a possible
quantum computation device.  After analyzing the change in the level
structure of a non-uniform confining potential as a periodic potential
is superimposed to it, we show how this structure combined with the
Pauli principle and fermion degeneracy can be exploited to create unit
occupancy of the lattice sites with very high efficiency.
\end{abstract}

\maketitle

Ultra-cold neutral atoms list among the most promising candidates for
quantum computing (QC) applications \cite{roadmap}.  Several possible
QC schemes using cold atoms have recently been proposed
\cite{interaction}, and various groups are presently working for their
experimental realization \cite{neutralatom_exp}.  The common starting
point is the creation of a quantum register, i.e. a scalable array of
well-characterized qubits \cite{DiVincenzoFdP}.  The qubits are
usually indentified with two internal states, typically hyperfine states,
of a single atom. The atom-qubits need to be spatially well separated, and
yet to be at a distance which allows to efficiently activate their
interaction.  These requirements are nearly ideally fulfilled by a
periodic potential with exactly one atom localized at each lattice
site. Such a periodic potential can be generated, for instance, by
counter-propagating laser beams or with microscopic magnetic surface
traps (atom chips).  As a natural consequence of the vast knowledge
developed in the last decade in the field of Bose-Einstein
condensation \cite{BEC}, until now most efforts towards atomic QC
implementations have been directed at the use of bosonic atoms.  Only
recently an increasing interest has been devoted to the cooling and
manipulation of fermionic atoms in confining potentials
\cite{fermions,fermionic-ketterle} and optical lattices
\cite{fermionic-lattices}. In particular the possibility to create a
quantum register with fermionic atoms in a homogeneous optical lattice
via coherent filtering has been proposed in \cite{daley03}.

In a uniform system, a lattice with a single atom per site can be
created by confining $N$ atoms in a one-dimensional box of length $L$,
and by subsequently superimposing a periodic potential of spatial
period $d$, so that the number of lattice sites in the box is $L/d=N$.
If the periodic potential is raised sufficiently slowly and if the
effective atom-atom interaction is repulsive, the system ends up in
its ground state which has an average occupation of one atom per
site. One could equivalently use bosons with a positive scattering
length or polarized fermions, in which case the Pauli principle
provides a strong effective repulsion. In either case, however, this
simple scheme has significant problems.  To avoid the creation of
defects like holes or multiply occupied wells, it requires equal
number of particles and lattice wells besides a long adiabaticity
time.

In this Letter we show that, in the presence of a non-uniform external
confinement instead, fermionic atoms have crucial advantages over
bosonic atoms and we propose a simple protocol to create a quantum
register which directly exploits the peculiar properties of the
Pauli principle.

Before discussing our proposal, it is useful to further consider the
case of non-interacting fermions confined by a homogeneous box.  In
the absence of the periodic potential, the $N$ fermions occupy the $N$
lowest single-particle levels up to the Fermi energy
$\epsilon_F=\pi^2\hbar^2N^2/2mL^2$, with $m$ being the atomic mass.
When the periodic potential is raised, in order for the number of
lattice sites to be equal to the number of particles we should have $
E_r=\epsilon_F$, where $E_r=\pi^2\hbar^2/2md^2$ is the recoil energy
of the lattice. Increasing the depth of the optical lattice slowly
enough to avoid excitations, the lowest $N$ levels gradually cluster
to form the first Bloch band, yet conserving their initial unit
occupancy due to the Pauli principle.  Eventually, in the limit $s\to
\infty$ the $N$ levels become degenerate and the Wannier
wavefunctions, which are $N$ linear combinations of the Bloch states
localized one in each well, are exact eigenstates of the system. Under
realistic conditions this procedure has the following drawbacks:

\noindent 1. For finite $s$, the Wannier states are only approximate
eigenstates and delocalize on a time scale of the order of
the inverse of the bandwidth $\delta$. This limits the time over which 
one can perform QC operations.

\noindent 2. Randomly distributed defects appear in the register

\noindent (i) if $\epsilon_F \neq E_r$ --- the condition
$\epsilon_F<E_r$ results in empty sites, while $\epsilon_F>E_r$
results in multiply occupied sites;

\noindent (ii) if the initial temperature is not much smaller than the
energy level spacing;

\noindent (iii) if the adiabaticity condition is not perfectly
fulfilled, i.e. if the raising time of the lattice is not large
as compared to the inverse level spacing.

The conditions required to create a lattice with unit filling
along the lines specified above are unrealistically strict, and
more complex schemes (see e.g. \cite{daley03, brennen}) are
needed in order to eliminate the defects produced in the register.

\begin{center}
\begin{figure}[h!]
\includegraphics[width=1\linewidth]{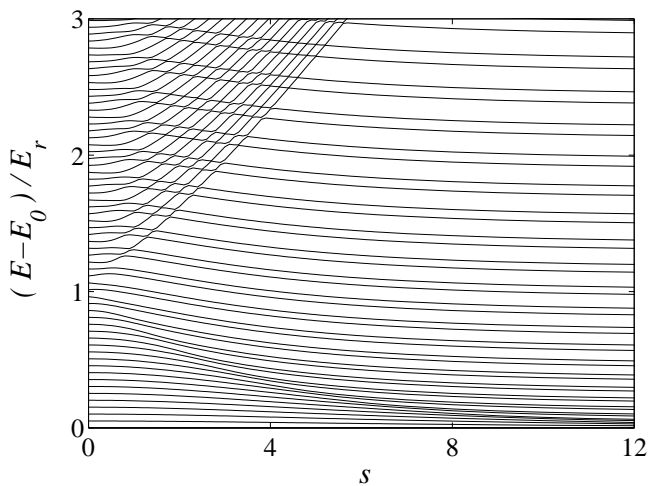}
\caption{Energy levels relative to the ground state $E_0$ as a
function of the optical potential depth $s$, for a harmonic
confinement with $\hbar \omega=E_r/(2 \pi^2)$ and $z_0=d/3$.} 
\label{fig_s}
\end{figure}
\end{center}

When the Pauli principle is combined with a non-uniform confinement,
the scenario changes completely.  We first consider a one-dimensional
geometry, which can be realized in practice with a tight radial
confinement, so as to freeze excitations of the radial degrees of
freedom. The extension to higher dimensions will be briefly discussed later.  
In the presence of an external
confinement, $V_{ext}$, the total trapping potential is \beqn
V(z)=V_{ext}(z)+sE_r\sin^2(\pi z/d), \eeqn where $s E_r$ is the
lattice height. In order to present the corresponding loading
procedure, we need to analyze the behavior of the energy eigenvalues
and the corresponding eigenstates as a function of the optical
potential depth. In Fig.~\ref{fig_s}, we consider the specific case of
a harmonic external confinement $V_{ext}(z)=m\omega_z^2 (z-z_0)^2/2$,
where $z_0$ is an arbitrary offset between the center of the confining
potential and the optical lattice.
As the periodic potential is raised, the eigenstates change from the
well known oscillator states at $s=0$, to states localized at the
lattice sites for large $s$.

It is a general feature that the eigenvalues of the initial external
confinement (\mbox{$s=0$}) can be grouped into two sets:

\noindent (I) The states with energy smaller than the recoil energy
change smoothly as $s$ increases;

\noindent (II) The levels with energy larger than the recoil energy
show a more complicated behavior characterized by several avoided
crossings.

\begin{center}
\begin{figure}[ht!]
\includegraphics[width=1\linewidth]{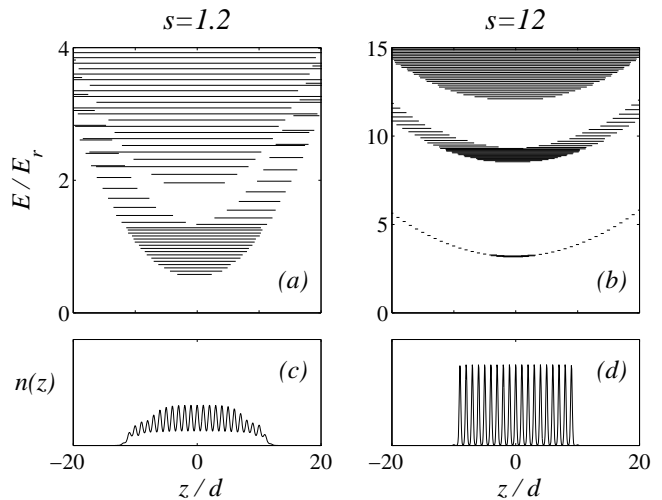}
\caption{Spatial position and extension of the eigenfunctions versus
the corresponding eigenvalues for $s=1.2$ (a), and $s=12$ (b); Density
profiles calculated for a number of atoms such that $\epsilon_F=E_r$
at $s=0$, for $s=1.2$ (c), and $s=12$ (d). The confining potential is
the same as in Fig.~\ref{fig_s}.}
\label{cipolla}
\end{figure}
\end{center}

The reason for this distinction, and the mentioned localization of the
eigenstates, can be understood by looking at the behaviour of the
eigenvalues and the eigenfunctions, illustrated in Fig.~\ref{cipolla}.
The figure shows the spatial position and the extension of the
eigenfunctions versus the corresponding eigenvalues for two values of
$s$.  When $s$ increases the eigenfunctions become more and more
localized in space and a band-like structures appear.  The gaps
between {\it pseudobands} open at the center of the trap, and the one
between the first and the second band separates the initial
eigenstates with energy smaller and larger than $E_r$, i.e. the states
in sets (I) and (II) above.  The eigenfunctions with initial energy
smaller than $E_r$ gradually localize, each one in a different well.
The energy of the eigenfunction in the well at position
$\ell d$ (with $\ell=0,\pm 1,\pm 2,\ldots$) is approximately given by
\begin{equation}
E_\ell\simeq \frac{1}{2} \hbar {\tilde \omega} + V_{ext}(\ell d),
\label{eig}
\end{equation}
where $\hbar\tilde\omega/2=\sqrt{s}E_r$ is the energy of the lowest
level of the lattice well. The first important observation is that,
contrary to the uniform system case, the localized states are here
exact eigenstates of the system for finite values of $s$. The amount
of localization of the eigenfunctions depends on $s$ and the external
potential. The second crucial point is that, by increasing $s$, states
with higher initial energy evolve into states localized farther away
from the origin as compared to those with lower initial energy. For
the register loading procedure, this implies that the constraint
$\epsilon_F=E_r$ of the homogeneous system is relaxed to
$\epsilon_F\leq E_r$, since the missing particles result in empty
holes localized at the wings, and not randomly distributed as
before. These facts change dramatically the loading conditions and are
the major results of the present work.

For large $s$, the energy spectrum (\ref{eig}) is characteristic of
all states in the first pseudoband, while the states in the second
pseudoband have energy approximately given by $3 \hbar
{\tilde \omega}/2 + V_{ext}(\ell d)$.  One could have hoped that the
condition $E_r\geq\epsilon_F$ could be further relaxed, since the
energy of the states in the second pseudoband progressively increases
for increasing $s$, leaving more and more localized states of the first
pseudoband energetically allowed. This is, in fact, not the case as
can be seen by looking at the behaviour of the levels of set
(II). These initially correspond to states in different
pseudobands. Due to the localization effect, as soon as a gap opens,
there is very little spatial overlap between eigenfunctions with
similar energy belonging to different pseudobands. The small spatial
overlap corresponds to a very small tunnelling between almost
degenerate states. This fact is also clearly reflected in the
structure of levels in Fig.~\ref{fig_s}, where for increasing $s$ the
gaps at the avoided crossings very quickly tend to disappear. For our
loading purposes this means that over the time scales of interest, the
atoms will remain trapped in their pseudoband even if, by increasing
$s$, states in lower pseudobands have much lower energy. Thus if
initially $\epsilon_F>E_r$, the ground state of the final potential
cannot be reached in realistic time scales and one will be left with
more than one particle in the central wells.

The level structure of the combined potential and the Pauli principle
have crucial consequences also on the effects of a finite temperature,
$T$, and on the adiabaticity time condition. We here summarize all the
properties in a detailed comparison with the uniform system case:

\noindent 1. For large but finite $s$ the eigenstates of the
Hamiltonian can be well localized, depending on the
particular shape of the external confinement.

\noindent 2.  An inner part of the register virtually free of defects
can be generated since:

(i) The strict requirement $\epsilon_F=E_r$, fixing the initial number
of atoms, is relaxed to $\epsilon_F<E_r$;

(ii) In an adiabatic process at finite $T$, the occupation number of
the final localized states is equal to that of the corresponding level
of the confining potential. For the states at lower energy in
current experiments this can differ from unity by as little as one part in
$10^{8}$ \cite{fermionic-ketterle}.  As long as $T\ll T_F$ and
$f(E>E_r)\ll 1$ (with $f$ being the Fermi thermal distribution), the
states with occupancy significantly different form one are localized
in a region $\sim k_BT$ around the Fermi level.

(iii) Strict adiabaticity is not required. Since the excitations
induced by a potential varying in a time of the order of $ \tau$ are
only created around the Fermi level in a shell of the order of
$\hbar/\tau$, one just has to enforce that the raising process be slow
compared to the inverse of the Fermi energy. For instance in the case
of a 1D harmonic oscillator, the condition becomes $\tau
\gg \hbar / \epsilon_F = 1/ ( N \omega)$, allowing a factor of $N$ faster
initialization than in the bosonic case.

We remark that the points (ii) and (iii) are direct consequences of
the fermionic nature of the atoms, and do not have a direct analogue
in the Mott insulator phase of trapped bosons \cite{mott}.

\begin{center}
\begin{figure}[h!]
\includegraphics[width=1\linewidth]{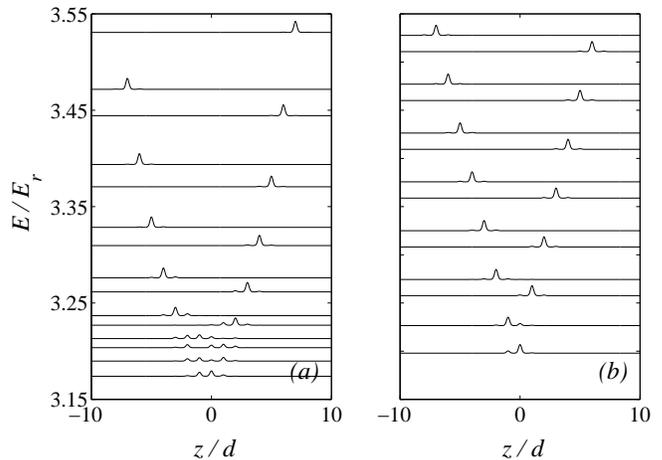}
\caption{Localization of the lowest eigenfunctions for $s=12$ for a
harmonic confinement as in Fig.~\ref{fig_s} and for a V-shaped
confinement $V=|z-z_0| E_r/(2 \pi^2 d) $ (b).  The V-shaped potential
is chosen to have the same slope as the harmonic one at $z = 4 d$.}
\label{loc}
\end{figure}
\end{center}

To complete our analysis of the loading it is necessary to discuss
more in detail the localization of the eigenstates. In
Fig.~\ref{cipolla}, we plot the density profiles at $T=0$ for two
values of $s$, with a number of atoms such that $\epsilon_F=E_r$. We
see that the density becomes flat in the presence of a deep periodic
potential \cite{pezze}. A uniform density distribution, however, does
not imply the perfect initialization of the quantum
register. Comparing Fig.~\ref{cipolla}(d) and Fig.~\ref{loc}(a)
(corresponding to the same parameters), we see that
even if the density distribution is flat, the lowest eigenfunctions
are not yet localized.


The degree of localization of the eigenfunctions at site $\ell$ arises
from the interplay between the tunneling $\delta$ and the energy
difference between neighboring wells at position $\ell d$,
given by $\Delta E_{\ell} \approx 
d \, [\, \partial V_{ext} (z) /  \partial z \,]_{z=\ell d}$.
In the large $s$ limit, the eigenstates will be localized in the
single wells, apart from a small contamination with first neighbors.
The probability amplitude for the atom localized at site $\ell$ to
find itself in the site $\ell \pm 1$ is given by $\alpha \approx \delta /
(2 \Delta E_{\ell})$.
This simple relation, whose validity was also checked numerically,
implies that it is the first derivative of the potential to be
important for localization. In the case of a harmonic potential
then, the atoms are very well localized in the outer wells, while the
central wells always suffer some tunnelling, even if small, due to the
vanishing derivative. In a V-shaped potential $V_{ext}(z)= \mu_M
\nabla B |z-z_0|$ (with $\mu_M$ the magnetic moment and $\nabla B$ a
constant magnetic field gradient), all wells are instead equivalent in
this respect, as shown in Fig.~\ref{loc}.

Consequently, while on the one hand in a shallow confining potential a
larger number of atoms can be loaded following the above scheme, since
more levels are found below the recoil energy, on the other hand, once
the lattice is raised, a steep potential is required to get a good
localization of the eigenstates.  A convenient procedure to obtain
both a large number of particles and a good localization is, thus, to
first raise the optical lattice up to its maximum depth in the shallow
trap (determined e.g. by $\omega_{in}$ or $\nabla B_{in}$), and then
to make the external confinement steeper (determined by $\omega_{fin}$
or $\nabla B_{fin}$) in order to ensure good localization of all
populated levels.

With a very steep final confinement, it might happen that states localized in
sites away from the center of the trap become higher in energy
than states in the second or higher pseudobands localized in the
center of the trap. This happens if $V_{ext}(\ell d) > 3 \sqrt{s}
E_r$.  This is not a problem as long as states in different
pseudobands are localized in different regions of space and the
tunneling between those states is negligible.  
The outermost atom in the register localized at $\ell_{max}=N/2$ is
degenerate with the state in the second pseudoband localized around
site $\ell'$, given by $V_{ext}(\ell' d)+\hbar{\tilde
\omega}=V_{ext}(\ell_{max} d)$. The condition for negligible
tunneling is $\ell_{max} -\ell' \gg 1$.  The explicit expressions for
harmonic and V-shaped confinement are respectively $\ell_{max} -\ell'
= (8/\pi^2) \sqrt{s} N (\omega_{in}/\omega_{fin})^2$ and $\ell_{max}
-\ell' = (8/3) \sqrt{s} N (\nabla B _{in}/ \nabla B_{fin})^2$.

We now consider, as specific examples, $^{40}$K atoms in a lattice
with periodicity $d=\lambda/2 = 431.5$ nm and $^{6}$Li atoms in a
lattice with periodicity $d=\lambda/2=335.5$ nm. For both species, we
take the magnetic moment equal to the Bohr magneton $\mu_B$.  The
first step of our procedure consists in raising the optical lattice to
its final depth.  The next step is to make the external confinement
tighter. The conditions fixing the strength of the final
confinement are a good localization of the eigenstates ($\alpha <
10^{-2}$) and a negligible tunneling to higher bands
($\ell_{max}-\ell' \gg 1$).  In Table I, we show the
parameters required to trap a maximum number of atoms
$N=E_r/\epsilon_F=10^3$.  The final optical lattice depth is $s=30$,
corresponding to a tunneling parameter $\delta=10^{-3} \,E_r$.
Lighter atoms require stronger confinements in
order to be well localized. Moreover, as already discussed in
Fig.~\ref{loc}, the V-shaped potential is more efficient than the
harmonic confinement, since the eigenstates
are localized to the same degree throughout the register.

\begin{widetext}

\begin{table}[h]
\begin{tabular}{|c||c|c|c|c|c||c|c|c|c||}
\hline
\begin{minipage}{1.5cm}
\begin{tabular}{c} \end{tabular}
\end{minipage}
 &  
\multicolumn{5}{c||}{Harmonic confinement} &  \multicolumn{4}{c||}{V-shaped confinement} \\
\hline
& $\omega_{in}$ & $\omega_{fin}$ & $\alpha_{N/20}$ & $\alpha_{N/2}$ & $\;\ell_{max}-\ell'\;$ 
               & $\nabla B_{in}$ & $\nabla B_{fin}$ & $\alpha$ &  $\;\ell_{max}-\ell'\;$ \\ \cline{2-10}
$^{40}$K & $2 \pi \times 6.7$ Hz & $2 \pi \times 170$ Hz&  $3 \times 10^{-2}$ &  $3 \times 10^{-4}$ & 7 
         & $0.8$ G/cm & $10$ G/cm & $5 \times 10^{-3}$ & $120$ \\
$^{6}$Li &  $\;2 \pi \times 74$ Hz & $\;2 \pi \times 1.9$ KHz $\;$ & $\;3 \times 10^{-2}\;$ & $\;3 \times 10^{-4}\;$ & 7
         & $\;1.2$ G/cm $\;$ & $\;140$ G/cm $\;$  &  $\;5 \times 10^{-3}\;$  & $120$ \\
\hline
\end{tabular}
\caption{Experimental parameters for the realization of a quantum
register with $^{40}$K and $^{6}$Li atoms, where we fix $N=10^3$ and
$s=30$.  The parameter $\alpha_{\ell}$ describes the localization
around site $\ell$ (which is site-dependent in the harmonic case).}
\end{table}

\end{widetext}

The extension of our proposal to 2D and 3D lattices is straightforward, as far as 
the relation $\epsilon_F<E_r$ fixing the maximum achievable number of atoms, the 
adiabaticity condition and the effect of finite temperature are concerned.
However, the localization properties of the eigenstates have to be investigated
carefully.
A higher dimensional register can contain a larger number of atoms. For
instance, in the case of cubic lattice and harmonic confinement, the
problem scales in a simple way with dimensions. In 3D, defining
$\bar a_{\rm ho}=\sqrt{\hbar/m\bar\omega}$ and
$\bar\omega=(\omega_x\omega_y\omega_z)^{1/3}$, we have $N \simeq (\pi^2\bar
a^2_{\rm ho}/2d^2)^3$, which can be easily of the order of $10^6$.

We acknowledge useful discussions with P. Zoller and H. Ott. This work has been
supported by the CRS-BEC INFM and by the European Commission under
contracts IST-2001-38863 (ACQP) and HPRN-CT-2000-00121 (QUEST),
co-financed by MIUR. LV acknowledges support from the University of Milan.



\begin{thebibliography}{99}



\bibitem{roadmap}
ARDA Quantum Information Science and Technology Roadmap, {\tt
http://qist.lanl.gov}.

\bibitem{interaction} G. K. Brennen, {\it et al.},
Phys. Rev. Lett. {\bf 82}, 1060 (1999); D. Jaksch, {\it et al.},
Phys. Rev. Lett. {\bf 82}, 1975 (1999); T. Calarco, {\it et al.},
Phys. Rev. A {\bf 61}, 22304 (2000); D. Jaksch, {\it et al.},
Phys. Rev. Lett. {\bf 85}, 2208 (2000); G. P. Berman, {\it et al.},
Phys. Rev. A {\bf 65}, 012321 (2002); G. K. Brennen, {\it et al.},
Phys. Rev. A {\bf 65}, 22313 (2002).

\bibitem{neutralatom_exp} O. Mandel, {\it et al.}, Phys. Rev. Lett.
{\bf 91}, 010407 (2003); R. Dumke, {\it et al.}, Phys. Rev. Lett. {\bf
89}, 097903 (2002)

\bibitem{DiVincenzoFdP}
D. P. DiVincenzo, Fort. Phys. {\bf 48}, 771 (2000).


\bibitem{BEC} C. J. Pethick and H. Smith,
{\it Bose-Einstein Condensation in Dilute Bose Gases},
Cambridge University Press 2002;
L. Piteavskii and S. Stringari, {\it Bose-Einstein Condensation},
Clarendon Press, Oxford, 2003.

\bibitem{fermions} K. M. O'Hara, {\it et al.}, Science {\bf 2179}
  (2002); G. Modugno, {\it et al.}, Science {\bf 297}, 2240 (2002);
  C. A. Regal,  {\it et al.}, Nature {\bf 424},
  47 (2003); S. Jochim, {\it et al.}, Science {\bf 302}
  5653 (2003); K. E. Strecker, {\it et al.},
  Phys. Rev. Lett. {\bf 91}, 080406 (2003); J. Cubizolles, {\it et al.},
  Phys. Rev. Lett. {\bf 91}, 240401 (2003);


\bibitem{fermionic-ketterle}
Z. Hadzibabic, {\it et al.},
Phys. Rev. Lett. {\bf 91}, 160401 (2003).

\bibitem{fermionic-lattices}
H. Ott, {\it et al.}
cond-mat/0311261;

G. Modugno, {\it et al.}, 
Phys. Rev. A {\bf 68}, 011601(R) (2003).

\bibitem{daley03}
P. Rabl, {\it et al.},
Phys. Rev. Lett. {\bf 91}, 110403 (2003).

\bibitem{brennen} G. K. Brennen, {\it et al.}, quant-ph/0312069.


\bibitem{mott} D. Jaksch, {\it et al.}, Phys. Rev. Lett. {\bf 81},
3108 (1998);
%
M. Greiner, {\it et al.}, Nature {\bf 415}, 39 (2002);
%
S. Peil, {\it et al.}, Phys. Rev. A {\bf 67}, 051603 (2003).

\bibitem{pezze} L. Pezz\`e, {\it et al.} (unpublished).

\end{thebibliography}
\end{document}